\documentclass[a4paper, runningheads]{llncs}

\newcommand{\orcidJM}	{\href{https://orcid.org/0000-0002-6332-5801}{\protect\includegraphics[scale=0.045]{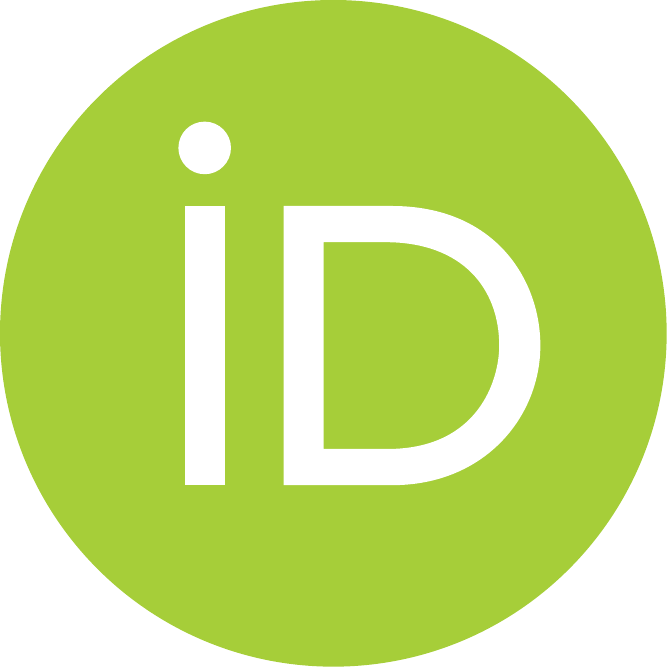}}}

\newcommand{\orcidSRM}	{\href{https://orcid.org/0000-0001-5656-6108}{\protect\includegraphics[scale=0.045]{orcid.pdf}}}

\usepackage[USenglish]{babel}
\addto\captionsUSenglish{\renewcommand{\figurename}{Fig.}}
\usepackage{lmodern}
\usepackage{adjustbox}
\usepackage{tabto}
\usepackage{booktabs}
\usepackage{tabularx}
\usepackage{makecell}
\usepackage{rotating}
\usepackage{graphicx}
\usepackage{tikz}
\usepackage{pgfplots}
\pgfplotsset{compat=1.15}
\usepackage{multicol,multirow}
\usepackage{amssymb,amsmath}
\usepackage[autostyle, english=american]{csquotes}
\usepackage{url}
\usepackage{xcolor}
\usepackage{placeins}
\usepackage[inline, shortlabels]{enumitem}
\usepackage{cite}
\usepackage{breakcites}
\usepackage{listings}

\usepackage[parfill]{parskip}

\newcommand{\ici}{\raisebox{-0.5pt}{\includegraphics[height=0.11in]{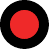}}}
\newcommand{\idsi}{\raisebox{-0.5pt}{\includegraphics[height=0.11in]{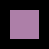}}}
\newcommand{\ioi}{\raisebox{-0.5pt}{\includegraphics[height=0.11in]{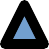}}}

\usepackage{todonotes}

\usepackage{xcolor}
\definecolor{lightgray}{rgb}{.9,.9,.9}
\definecolor{darkgray}{rgb}{.4,.4,.4}
\definecolor{forestGreen}{RGB}{34,139,34}
\definecolor{orangeRed}{RGB}{255,69,0}
\definecolor{codegreen}{rgb}{0,0.6,0}
\definecolor{codegray}{rgb}{0.5,0.5,0.5}
\definecolor{codepurple}{rgb}{0.58,0,0.82}

\RequirePackage[bookmarks=true,
                urlcolor=black,
                citecolor=black,
                linkcolor=black,
                pagecolor=black,
                bookmarksdepth=3,
                bookmarksopen=true,
                colorlinks,
                hyperfigures]{hyperref}

\makeatletter
\def\RemoveSpaces#1{\zap@space#1 \@empty}
\makeatother

\makeatletter
\newcommand{\printfnsymbol}[1]{%
  \textsuperscript{\@fnsymbol{#1}}%
}
\makeatother

\begin{document}

\title{Cloud Process Execution Engine: Architecture and Interfaces}

\titlerunning{CPEE}

\author{Juergen Mangler\orcidJM \and
Stefanie Rinderle-Ma\orcidSRM}

\authorrunning{J. Mangler et al.}

\institute{Department of Informatics, Technical University of Munich, \\85748 Garching, Germany
\\ \email{\{firstname.lastname\}@tum.de}}

\maketitle
\begin{abstract}

Process Execution Engines are a vital part of Business Process Management (BPM)
and Manufacturing Orchestration Management (MOM), as they allow the business or
manufacturing logic (expressed in a graphical notation such as BPMN) to be
executed. This execution drives and supervises all interactions between humans,
machines, software, and the environment. If done right, this will lead to a
highly flexible, low-code, and easy to maintain solution, that allows for
ad-hoc changes and functional evolution, as well as delivering a wealth of data
for data-science applications.

The Cloud Process Execution Engine CPEE.org implements a radically distributed
scale-out architecture, together with a minimal set of interfaces, to allow for
the simplest possible integration with existing services, machines, and
existing data-analysis tools.

Its open-source components can serve as a blueprint for future development of
commercial solutions, and serves as a proven testbed for academic research,
teaching, and industrial application since 2008.

In this paper we present the architecture, interfaces that make CPEE.org
possible, as well as discuss different lifecycle models utilized during
execution to provide overarching support for a wide range of data-analysis
tasks.

\end{abstract}

\setlist[enumerate]{align=left,leftmargin=2.3em}

\section{Introduction}
\label{sec:intro}

The Cloud Process Execution Engine is an open-source bare-bones radically
service-oriented process engine, that, together with a set of components forms
a Business Process Management (BPM) system that proved (a) great for teaching
as all internal mechanisms are exposed as REST
interfaces~\cite{mangler_quo_2009,mangler_origin_2010} and can be inspected,
used and augmented by interested students, and proved (b) great for research to
either experiment with the currently prevalent graphical modelling language -
Business Process Management notation (BPMN) - through extensions, or by
developing altogether novel modelling languages~\cite{mangler_cloud_2010}.
Implementing worklists, correlators, run-time data-analysis, self-healing
processes, or novel means of inter-process and inter-instance synchronization
is easy and stream-lined: external REST-services utilizing your
language/framework of choice do it all. Finally, (c) also companies took an
interest, due to scalable highly flexible architecture that scales from a
raspberry-pi with some instances to mainframes with 1000s of parallel running
processes, while efficiently utilizing multi-core architectures.

CPEE.org tries to further the low-code and model-based process execution
paradigm, that allows non-programmers to connect software, machines, and humans
in simple and easy to understand ways. By allowing for ad-hoc instance changes
to realize repair, as well as providing tools for process model versioning and
evolution, it wants to show-case features that will hopefully make it into many
current and future BPMs.

\section{BPM Basics}
\label{sec:basics}

Since many years, moving infrastructure and with it software components to the
cloud is an important topic when dealing with digitization. Business Process
Management is about graphical models containing sequences of activities,
decisions and parallel branches. Activities describe how to invoke (external)
functionality implementing an activity, including the required input, and how
to transform the expected output to be usable for subsequent activities and
decisions. Business Process Management (BPM) traditionally has been relying on
monolithic Process Engines, mostly written in Java, conceived in the late 90s,
and not changed much since then. The typically consist of the following tightly
coupled components as also depicted in Fig. \ref{fig:arch}:

\begin{figure}
  \centering
  \includegraphics[width=1.0\textwidth]{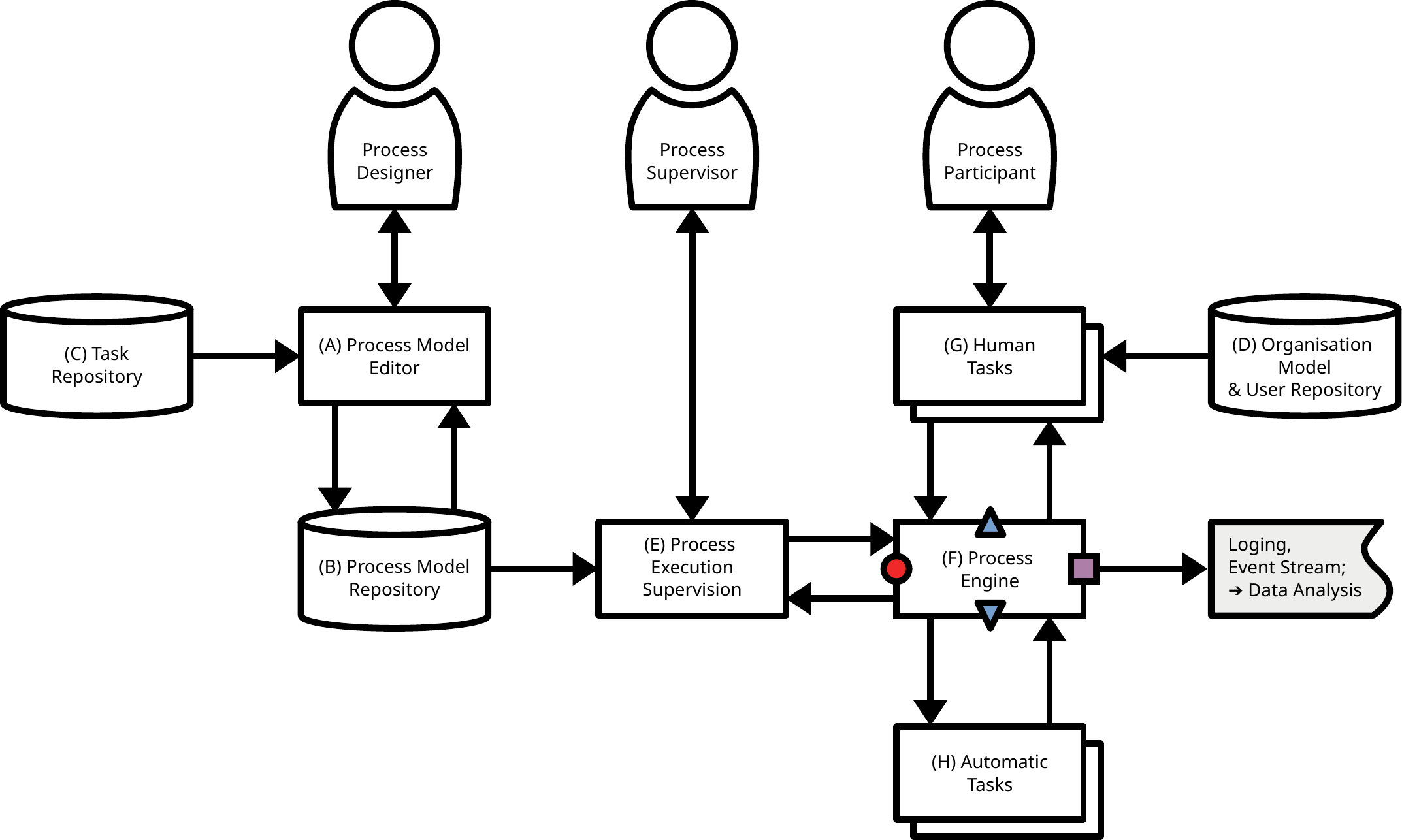}
  \caption{Architecture \& Stakeholders}
  \label{fig:arch}
\end{figure}

\begin{enumerate}

  \item[(A)] Process Model Editor (UI, nowadays probably BPMN 2.x, CMMN, DMN)

  \item[(B)] Process Model Repository

  \item[(C)] Task Repository

  \item[(D)] Organization Model and User Repository

  \item[(E)] Process Execution Supervision (UI)

  \item[(F)] Process Engine

  \item[(G)] Worklists, Dashboards: Human Tasks (UIs)

  \item[(H)] Invoked Applications: Automatic Tasks

\end{enumerate}

Thus the traditional stakeholders in a BPM system are the

\begin{itemize}

  \item \textbf{Process Designers:} they create the process at design time, and improve
  / evolve process models if necessary.

  \item \textbf{Process Supervisors:} instantiate processes and supervise their
  execution. They user

  \item \textbf{Process Participants:} take care of the work as modelled by human /
  manual tasks. They user their own independent UI, and potentially know
  nothing about the existence of a BPM.

\end{itemize}

The Process Model Editor \textbf{(A)} allows Process Designer to create and change
Process Models (PMs), which are stored in a Process Model Repository \textbf{(B)}.

A PM is not executable, until each activity is assigned the corresponding
functionality, and the required input/output parameters are set. The same goes
for events. For example a timer event requires additional information how long
to wait in machine-readable form. Typically the Task Repository \textbf{(C)}
holds a list of functionalities available to the Process Designer for
association with activities. This is true for automatic \textbf{(H)} as well as
human tasks \textbf{(G)}.

Human tasks \textbf{(G)} are typically come in the form of UIs called worklists
or dashboards (see Sec. \ref{sec:comp}, ``Components''), worklists require
information about their users for work assignment. When worklists target
work-distribution in organizations, user/role relationships are typically
utilized to automatically distribute work between all users of a role. If
worklists target customers, all work is assigned to one customer. In both cases
users have to be logged-in/identified. The same goes for dash-boards if they
allow for interaction, but it might also possible that interactions are
possible without being logged in, because of the assumption that only eligible
users have physical access to the dashboard. All information about users and
roles (Organizational Structure) is kept in a Organization Model and User
Repository \textbf{(D)}.

The Process Execution Supervision (UI) \textbf{(E)} is used to deploy PMs from
the Process Model Repository to the Process Engine \textbf{(F)} - the is used
to \textbf{create an instance (instantiate)}. Each instance is a \textbf{COPY} of the
model, changes to the instance might be possible (depending on engine features,
i.e. run-time adaptation), in which case the model of the instance might
deviate from its parent.

The Process Engine (PE) \textbf{(F)}, is in charge of executing the instance model, and
realizing all the invocations of external services or functionalities which are
represented by activities, events or gateways (furthermore referred to as
activity enactment). Process Engines most commonly are interpreters - just like
for example the Java Virtual Machine (HotSpot). Other engines like CPEE.org are
transpiling the model to lower-level languages for compilation/execution.

Invoked applications \textbf{(H)} are either typically either realized as Java
components, that can be loaded into the process engine, or as external
services, that have to implement a certain API as provided by the BPM software
provider (e.g. Camunda provides APIs for Java, and JavaScript, unofficial
Python support exists\footnote{Last checked: 2022-09-25}). Depending on the PE
these services might either be realized as REST/SOAP/OPC-UA/...

\section{Process Engine Interfaces}
\label{sec:interfaces}

\begin{figure}
  \centering
  \includegraphics[width=1.0\textwidth]{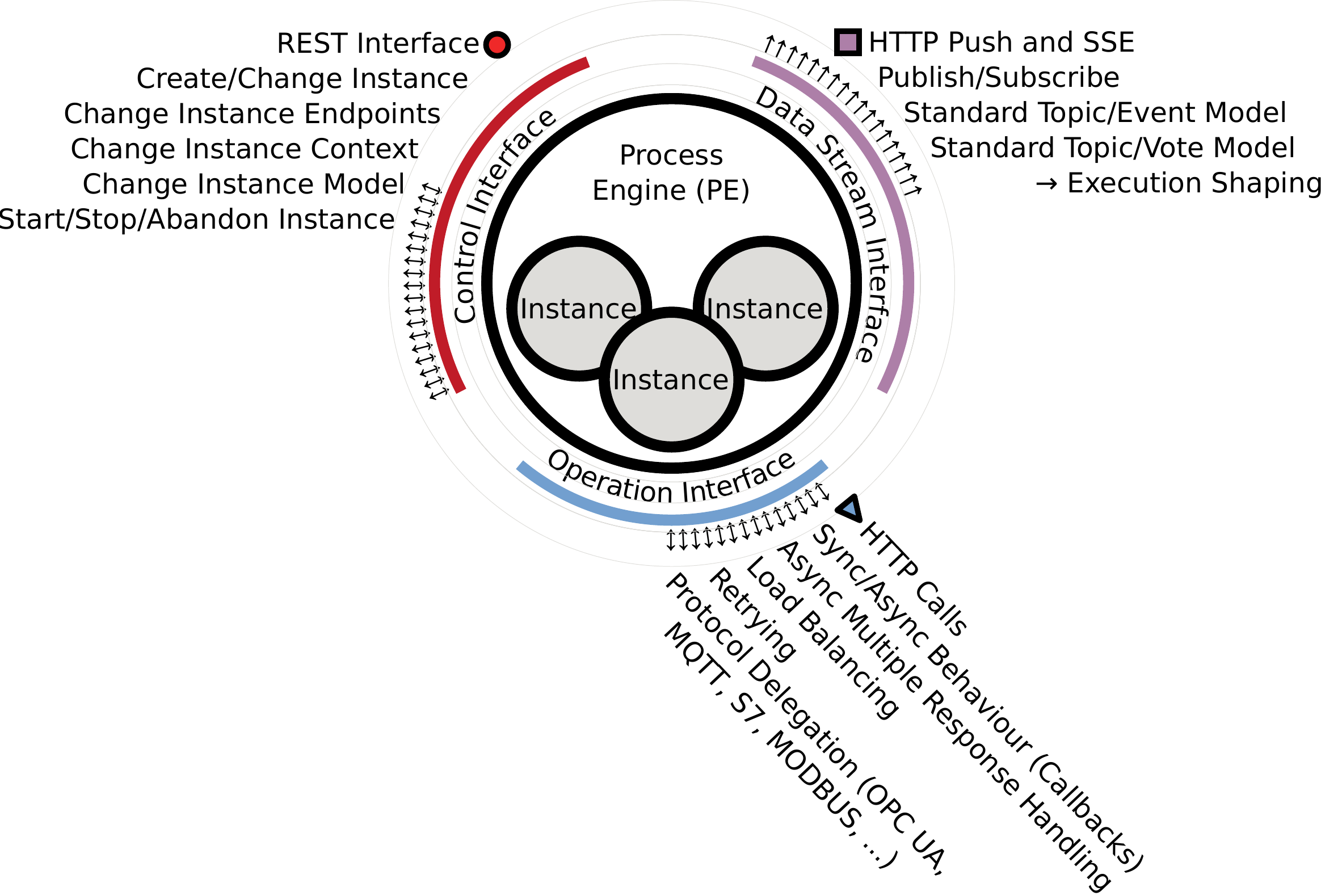}
  \caption{Interfaces of CPEE.org}
  \label{fig:cpee-arch1}
\end{figure}

The \textbf{PROCESS ENGINE (PE)} is at the heart of the architecture. All other
components contribute data that is required for the execution of an instance,
but the process engine executes the process and coordinates the interaction of
all components, no matter if they are standardized (\textbf{(E)}, \textbf{(G)})
or user-created (\textbf{(G)}, \textbf{(H)}).

While the architecture given in Fig. \ref{fig:arch}, might be either monolithic
or partially monolithic (components may be part of one big software package,
either desktop-based or web-based), it is also possible to separate all
components through clearly defined interfaces. The architecture of CPEE.org
takes the second approach, and is realize as a set of loosely coupled services.
Note that this is not necessary for the \textbf{(A)} Process Model editor, or
any other editor for that matter, as e.g. a Task Editor only contributes to the
\textbf{(C)} Task Repository, and an Organization Editor only contributes to
\textbf{(D)} Organization Model \& User Repository. So all editors interface if
the engine through the data structures they produce ex-ante. These
data-structures might be BPMN 2.0 Interchange format or CPEE.org trees for models,
custom LDAP structures for Organization Models \& User Lists, and proprietary
lists for tasks.

So at runtime, the process engine really has three interfaces with active
components, as marked through \ici{} \ioi{} \idsi{} as depicted in Fig.
\ref{fig:arch} and Fig. \ref{fig:cpee-arch1}.

\subsection{Control Interface}
\label{sec:ci}

\ici{} The Control Interface allows UIs used by the Process Supervisor to
perform all operations related to creating, starting and managing process
instances.

In order to \textbf{create an instance}, a Process Execution
Supervision (PES) UI has to allow to, (a) select a process model, (b)
instantiate it, (c) and supervise its execution through an execution engine:

\begin{itemize}

  \item Which activities are currently enacted/running?

  \item What is the current process context (the data-elements that exist
  during execution)?

  \item Which sub-processes have been spawned by the instance?

\end{itemize}

CPEE.org is an adaptive process engine, thus when an instance fails (stops),
through the PES the following operations can be performed:

\begin{itemize}

  \item Change endpoints for activities, i.e. change the functionality an
  activity invokes during enactment. E.g., when one production machine fails a
  second one might take over a task.

  \item Changing the process context, i.e. activities might yield faulty data
  that would prevent the successful execution of their instance. Manual changes
  to data-elements might save the instance.

  \item Change the thread of control, i.e. change which activities are enacted
  next. This might include skipping activities, but also re-doing activities.
  While re-enacting activities, or skipping activities might be harmful, as tasks
  typically have consequences, it might also be possible that
  re-enacting/skipping might save the instance. Thus any process engine should
  allow a Process Supervisor to do both.

  \item Re-starting the instance execution if it stops. As functionalities
  implementing activities can be temporary down, or endpoints/dataelements/thread
  of control can be changed to alleviate problems, re-starting the execution at a
  certain point is beneficial.

  \item Change the instance model, i.e. whenever a process instance is stopped,
  it might be necessary to change the instance model (e.g. repair the instance by
  inserting or deleting activities). Changed instance model become singletons -
  the no longer are identical to the process model they have been initially
  instantiated from.

\end{itemize}

CPEE.org strictly relies on a REST-interface to achieve all these changes.

\subsection{Operation Interface}
\label{sec:oi}

\ioi{} The purpose of the operation interface is to delegate and monitor the work
described by activities. Activities (A) are modelling elements in BPMN (or
other graphical notations), that describe how to invoke (external)
functionality (F), including the required input, and how to transform the
expected output to be usable for subsequent activities and decisions. By
default the functionality is assumed to be a black box, from the point of view
of a process engine (PE) it is not important what is going on inside. In other
words: the PE manages the data-flow to and from these activities.

\begin{figure}[!h]
  \centering
  \includegraphics[width=0.8\textwidth]{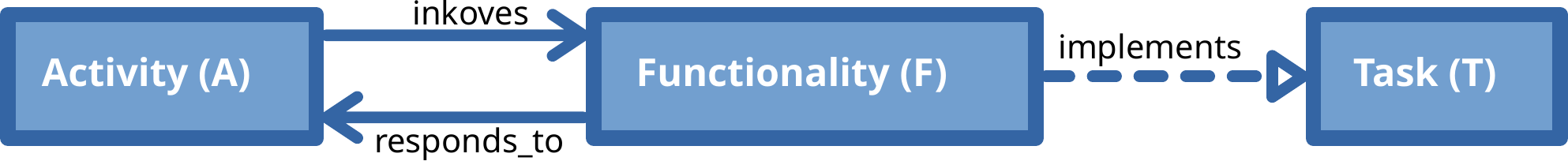}
  \label{fig:tf}
\end{figure}

Many process engines rely on an API to implement the functionality invoked
activities, i.e., F is implemented using an API, which allows to either (1)
load F into the PE (old-school monolithic engines), or (2) start F as a server
which can than be invoked by the PE enacting an activity. This has the
advantage that the protocol utilized between A and F is not important and can
be changed without affecting the implementation of F.

Alternatively, CPEE.org and other engines rely on protocol extensions instead
of an API to implement functionalities. CPEE.org‘s primary protocol
implementation is HTTP, which is extended by a set of CPEE.org-specific HTTP
headers allow for some special interaction patterns between PE and F:

\begin{figure}[!h]
  \centering
  \includegraphics[width=0.8\textwidth]{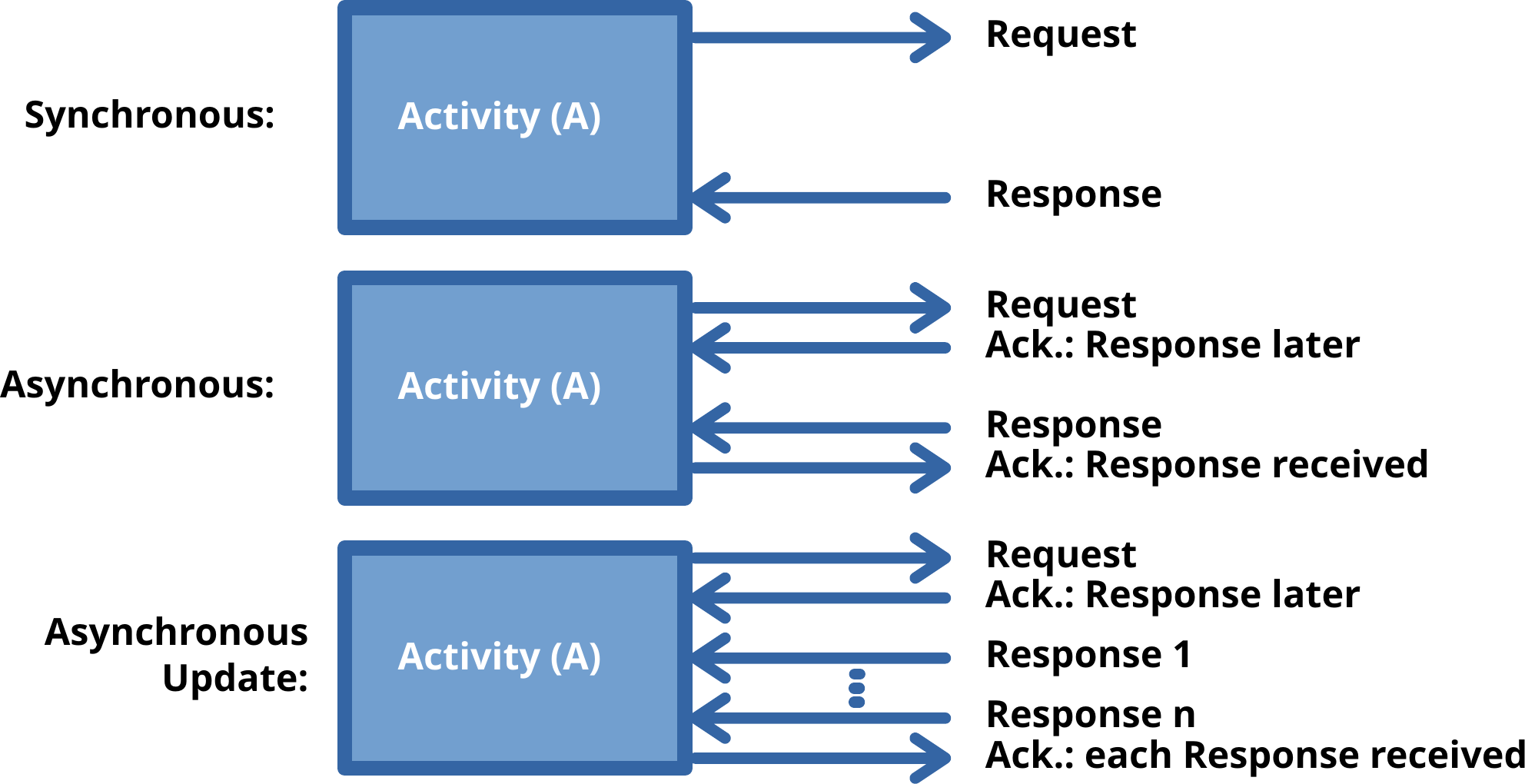}
  \label{fig:tf}
\end{figure}

For the \textbf{synchronous pattern}, answers are returned immediately, which in HTTP is
only possible if the answer is returned in a certain time-frame (about 30
seconds for normal network infrastructure). If the time-frame can not be
satisfied by F, the PE upon enactment of the activity in a certain instance
will receive a timeout, and the instance will be thus stopped. This type of
interaction can therefore only be used for simple and fast interactions.

The \textbf{asynchronous pattern} describes that (F) can delay the answer for as long as
necessary. This is done by telling the PE that the answer will arrive later.
This only works if a callback address is available to F. The PE maintains the
list of callback addresses. Each callback address allows the PE to forward the
answer to a certain activity in a certain instance.

Finally, the \textbf{asynchronous update pattern}, describes a special case of the
asynchronous pattern which allows to call back multiple times, e.g., to return
a series of status updates to the activity, or an arbitrary number of data
chunks (which is especially useful if large amounts of response data have to be
handled by T), This works by adding a flag to each answer, telling the PE if
further answers are to be expected.

All this in enabled by the addition of a minimal set of CPEE.org specific HTTP
headers, as enumerated below. Common CPEE.org HTTP headers, sent with each
request are:

\begin{itemize}

  \item \textbf{CPEE-BASE} - base location of the engine where the instance is running on\\
  (e.g., https://cpee.org/flow/engine/)

  \item \textbf{CPEE-INSTANCE} - instance number\\
  (e.g., 123)

  \item \textbf{CPEE-INSTANCE-URL} - url pointing to the instance\\
  (e.g., https://cpee.org/flow/engine/123)

  \item \textbf{CPEE-INSTANCE-UUID} - unique identifier of the instance\\
  (e.g., 059a4f32-dcb1-4ad0-a700-ddd3d1fbf64f)

  \item \textbf{CPEE-CALLBACK} - url to send any information to, should the implementation decide to answer asynchronously\\
  (e.g., https://cpee.org/flow/engine/123/callbacks/f8c24f12-1419)

  \item \textbf{CPEE-CALLBACK-ID} - unique identifier for the answer
  (e.g., f8c24f12-1419)

  \item \textbf{CPEE-ACTIVITY} - id of the activity invoking a functionality (e.g. a1)

  \item \textbf{CPEE-LABEL} - label of the activity invoking a functionality (e.g. Query Production Schedule)

\end{itemize}

Each response (independent of the pattern) can sent the following optional headers:

\begin{itemize}

  \item \textbf{CPEE-SALVAGE} - F communicates that it can currently not provide any
  answer, but might be available again later. This can be utilized by the PE in a
  fail-over scenario, to reroute the request to a different F or to retry the
  original F at a later point in time. If this header is present, its value is
  expected to be always ``true''.

  \item \textbf{CPEE-INSTANTIATION} - F communicates that it has instantiated a
  (sub-) process. F will most probably additionally return the instance-url in
  the body of the response (e.g., https://host2.cpee.org/flow/engine/124/). If
  this header is present, its value is expected to be always ``true''.

  \item \textbf{CPEE-EVENT} - F communicates that it a functionality-custom event should
  be included in the data sent out trough the \idsi{} data stream interface. This is
  especially useful if F has an internal lifecycle (e.g., if F implements a
  worklist) and wants to signal custom lifecycle transitions (such as a user
  taking or giving back a task). If this header is present, its value is expected
  to carry the name of the custom signal (e.g., worklist/task-taken).

\end{itemize}

The \textbf{asynchronous pattern}, in addition to the three optional common response
headers, has to use the \textbf{CPEE-CALLBACK} header with the ``Ack.: Response later''
message. If this header is present, its value is expected to be always ``true'',
and the PE will not continue the execution of the instance, but instead wait
for a reply. Each HTTP PUT to the \textbf{CALLBACK-URL} will prompt the PE to forward
the response to the activity and subsequently continue the instance.

The \textbf{asynchronous update pattern}, in addition to the three optional
common response headers, has to use the \textbf{CPEE-CALLBACK}, exactly the
same as the \textbf{asynchronous pattern}. For each response, additionally the
\textbf{CPEE-UPDATE} header is to used. Whenever this header is present and its
value is true, the PE forwards the message to a certain activity in a certain
instance, but the instance is not allowed to continue, and the activity
continues waiting for further responses. A response missing the
\textbf{CPEE-UPDATE} header is considered the last response, thus the PE will
forward the response to the activity and subsequently continue the instance.

Through these simple protocol extension, CPEE.org can support arbitrary
interactions. Custom protocols, such as OPC-UA (i.e., machine interfaces), can
be implemented as proxy F‘s. While the communication between A and F utilizes
the mechanisms described above, F will communicate with third-party services
and machines through custom protocol implementations.

\subsection{Data Stream Interface}
\label{sec:ds}

\idsi{} The purpose of the data stream interface is for a process engine PE to
communicate the state of \textbf{instances (I)} as well as the state of
\textbf{activities (T)} to micro-services connected to the interface.

The interface supports two ways a state communication: HTTP push to dedicated
URLs, and HTTP server sent events (SSE) upon request. In order to communicate
the state to external services, a PUB/SUB mechanism exists, that allows to
subscribe to a certain sub-set of events. Each subscription has to carry:

\begin{itemize}

  \item An optional endpoint, that denotes where to push the events. If the
  endpoint is omitted, it is assumed that the messages grouped by the
  subscription will be sent through SSE\footnote{SSE is actual a special kind of pull,
  where a client initiates a connection with a server, the server keeps the
  connection alive by heart-beating to the client, and can thus push data
  through the connection when it is available. It is thus a form of
  long-polling.} upon explicit request.

  \item An set of topics and events, that describe certain aspects of the
  execution of an instance.

\end{itemize}

Execution aspects, furthermore called topics, are:

\begin{itemize}

  \item The \textbf{state} topic contains a set of events describing potential instance
  states (see Sec. \ref{sec:il}).

  \item The \textbf{activity} topic contains a set of events describing the state of an
  activity in a particular instance (see Sec. \ref{sec:al}). Each activity is
  represented by a series of events, as an activity at least starts and finishes.
  As an instance might contain activities that are enacted in parallel, each
  event has to carry the activity id (e.g., t1). When an activity is enacted in a
  loop, the activity id is not enough to identify events belonging together, as
  through the network-based nature of event dispersal events might be
  out-of-order. Thus each enactment of an activity has to carry (in addition to
  the activity id) a unique enactment identifier, e.g., ``t1-enactment-1'' or
  ``t1-enactment-2''.

  \item The \textbf{position} topic allows to monitor the progression between activities.
  This includes events when activities become active, activities are no longer
  active, as well as events detailing the transition between two activities. A
  transition between two activities does not mean that they are in sequence, a
  transition might occur between an activity and a next activity based on a
  decision, or multiple activities might become active due to a parallel split.

  \item The \textbf{status} topic allows to monitor information about semantic execution
  properties of an instance, e.g., if a instance currently runs normally, or if
  some exception handling logic is active. The instance status can be changed as
  the result of any activity enactment.

  \item The \textbf{dataelements} topic allows to monitor the data-flow, independently of
  the control-flow of an instance. While the enactment of activities might change
  the process context (dataelements, variables), not each activity does so. Each
  event includes information about added, deleted and changed (from value, to
  value) dataelements.

  \item The \textbf{description} topic allows to monitor changes to the instance model.
  When ever an instance \textbf{is not running} (e.g., before an instance is started or
  when a instance stopped due to an error), changes/repairs to the process model
  might be applied. Changes can include assigning different functionality to an
  activity, inserting or deleting activities.

  \item The \textbf{endpoints} topic allows to monitor when a process instance links to
  new functionality. CPEE.org for each instance manages a list endpoints
  key/value pairs where each functionality is referenced by a key, e.g. timeout \textrightarrow
  https://cpee.org/functionalities/timeout/. Functionality is assigned to
  activities by this key. Change events can occur both at runtime (while an
  instance is executed) as well as while an instance is stopped. An activity
  might as part of their enactment dynamically change/adapt the endpoint list,
  namely changing the value of any key, resulting in activities invoking
  different functionality. This can be utilized to, for example, implement
  load-balancing or load-distribution. Each event includes information about
  added, deleted and changed (form, to) endpoints.

  \item The \textbf{attributes} topic allows to monitor changes to an instances
  attributes. Attributes might include the UUID of the process model an instance
  was originally spawned from (although the model of the instance might have
  changed), or an arbitrary number of labels and information assigned to the
  instance (name, author of the model, user responsible for repair, ...). Each
  event includes information about added, deleted and changed (form, to)
  attributes.

  \item The \textbf{condition} topic allows to monitor any decisions taken during the
  execution of an instance. This might include decisions taken based on xor/or
  gateways or loops. The event includes the condition, all involved dataelements
  and their values, as well as the result of the evaluation (true or false).

  \item The \textbf{task} topic groups a set of special and user-defined events, such as
  task/instantiation which is sent by functionalities implementing the creation
  instantiation of sub-processes instances. Such events are sent by
  functionalities trough the \ioi{} operation interface (see Sec.\ref{sec:oi}) and are
  subsequently distributed to subscribed services through the task topic. This
  interface is particularly useful for communicating the lifecycle or application
  state of functionalities trough a PE. For example, worklists/tasklists are just
  ordinary functionalities invoked as part of activity enactment. Although
  potentially a black box to the PE, worklists might have a fine-grained internal
  lifecycle dealing with how work is assigned to users, work on by users,
  including dealings with deadlines and conflicts. For runtime or ex-post
  data-analysis~\cite{stertz_analyzing_2020,stertz_temporal_2020} it van be very useful to include this information in subscribable
  data streams dispersed by the PE.

\end{itemize}

By selecting from this brad menu of topics external services can analyse all
aspects of an instance execution, regarding both data-flow and control-flow.
Each subscriber can use the data to, for example, write fine-grained execution
logs which include information far beyond the aspects specified in standards
such as the XES\footnote{\url{https://xes-standard.org/}} standard.

In addition to distributing events the interface also supports \textbf{execution
shaping}~\cite{mangler_cpee_2014}, which allows external services
subscribed to events to influence the execution, without invoking the \ici{} control
interface, but as part of the subscription to the \idsi{} data stream interface.

While normal events are sent by the PE without waiting or acknowledging a
response (fire and forget), special events, furthermore called votes are
treated differently. The PE waits for a response from each subscribed external
service and acts upon the responses. A minimal set of responses currently
implemented by CPEE.org includes:

\begin{itemize}

  \item \textbf{ack}: don‘t care or approval. Instance might continue to be executed as
  per the model.

  \item \textbf{callback}: answer will be sent later. Instance will remain in state
  running, but the activity referenced by the vote will remain frozen until the
  answer is received.

  \item \textbf{skip}: instance will remain in state running, the activity referenced by
  the vote will be skipped.

  \item \textbf{stop}: instance is stopped immediately.

  \item \textbf{start}: instance is started immediately.

  \item \textbf{value}: (1) the condition referenced in the vote is evaluating to the
  value (true, false). (2) the dataelement, endpoint, attribute referenced in
  the vote is set to the values.

\end{itemize}

Callback is the special case, that just delays the decision. All other
responses have to be unambiguous, with ack being the neutral response.
Examples

\begin{itemize}

  \item If 1..n services send skip and the rest of the services send ack, skip
  goes into effect.

  \item The same rule is applicable to stop and start.

  \item If 1..n services respond with action, but disagree on true/false, the
  rest of the services send ack, then the instance will be stopped.

  \item If 1..n services respond with value, but disagree on the actual value,
  the rest of the services send ack, then the instance will be stopped.

  \item If 1..n services respond with the same value, the rest of the services
  send ack, then the value will be set and the instance resume executing.

  \item Responses of value, skip, start/stop can be partially combined:

  \begin{itemize}

    \item value and skip can be combined, with value being enacted first, then
    the skipping the activity.

    \item value and start/stop can be combined with value being enacted first,
    then starting/stopping the instance.

    \item skip and stop can be combined, first the skipping the activity then
    stopping the instance.

    \item start and skip can be combined, first starting the instance, the
    skipping the active activity immediately.

    \item start, value, skip can be combined according to the schema above.

    \item value, skip, stop can be combined according to the schema above.

  \end{itemize}

  \item Dissenting start/stop responses can not be combined, the current state
  will remain.

\end{itemize}

Topics that have votes include:

\begin{itemize}

  \item \textbf{state} topic: start / stop can be prohibited or allowed. This is useful
  when implementing model checking techniques. Furthermore external services can
  change endpoints, attributes and dataelements on start or stop through the
  value response.

  \item \textbf{dataelements}, \textbf{endpoints}, \textbf{attributes} topics: changes to individual can be
  blocked (action) or corrected (value) . Furthermore, an execution of an
  instance can be stopped (stop response) in compliance checking scenarios.

  \item \textbf{condition} topic: the evaluation of conditions can be modified with the
  value response. Again the instance can be stopped (stop response) if necessary.

  \item \textbf{description} topic: Individual changes to the model can be prohibited
  through action responses.

\end{itemize}

With this powerful voting mechanism runtime conformance and compliance
checking, as well as self-healing, which all require not only certain data, but
also a set of actions to influence the execution, can be implemented through
external services.

The alternative would be, to allow external services to utilize the \ici{} control
interface, which would entail to always stop instances before changes, in order
to avoid race conditions. CPEE.org can thus cover the most important areas of
runtime process mining (discovery can be ignored in this context) and adaptive
process execution.

\section{Lifecycles}
\label{sec:lifecycles}

In order give a more detailed introduction to the \textbf{state}, \textbf{activity} and \textbf{task}
topics introduced as part of the \idsi{} data stream interface in the previous section,
this section will discuss the lifecycle models for:

\begin{enumerate}

  \item[4.1:] The execution of instances.

  \item[4.2:] The enactment of activities.

  \item[4.3:] The internal behaviour of worklist functionalities.

\end{enumerate}

\subsection{Instance Lifecycle}
\label{sec:il}

\begin{figure}
  \centering
  \includegraphics[width=0.5\textwidth]{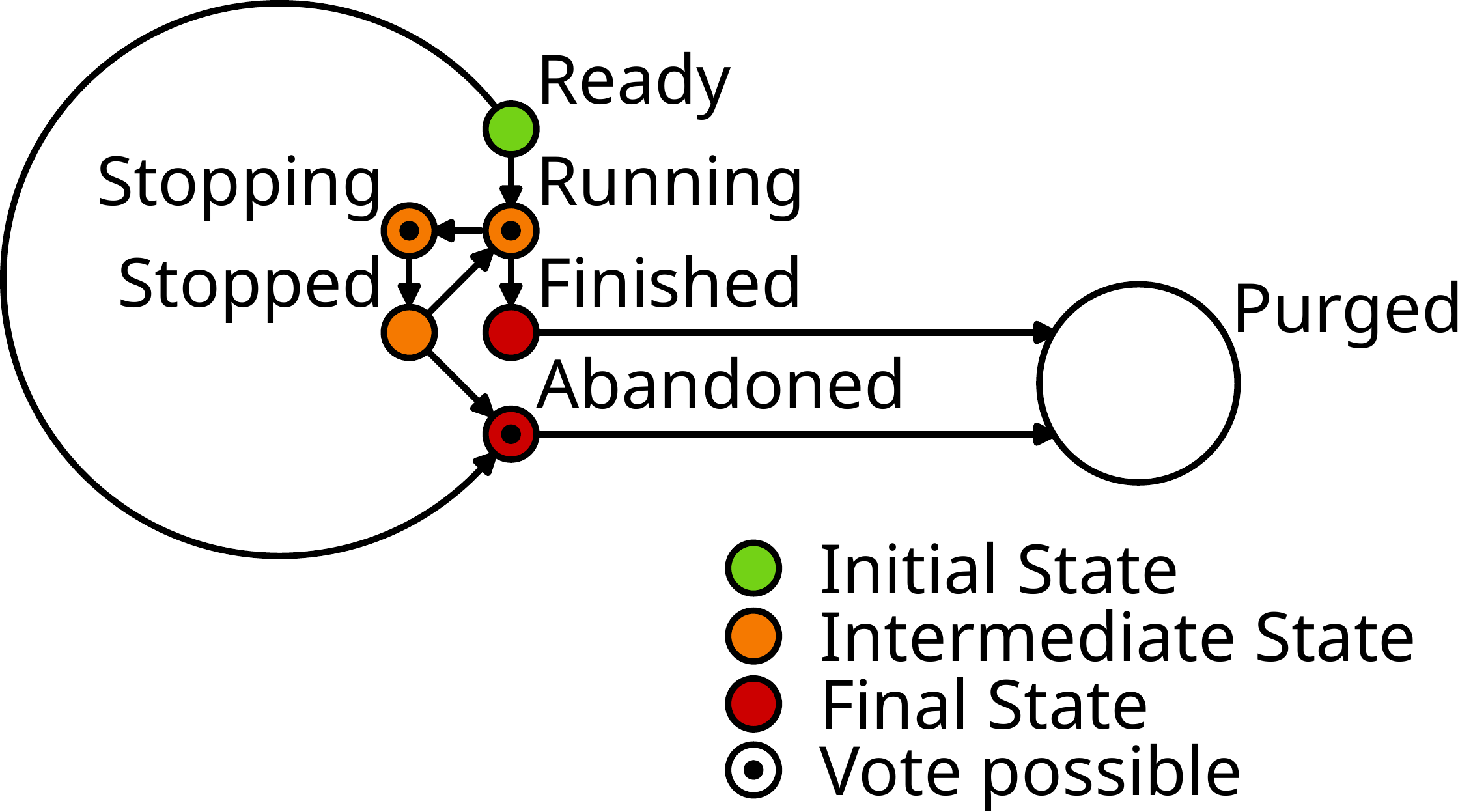}
  \caption{CPEE.org Instance Execution Lifecycle}
  \label{fig:lifecycle-instance}
\end{figure}

While the PE executes an instance, it goes through a number of states (see
Fig. \ref{fig:lifecycle-instance}). Reaching a state also results in sending an event through the \idsi{} data
stream interface for all external services subscribed to the \textbf{state} topic.

\textbf{Ready} is the state that an instance is in, immediately after it is created.
Instances in CPEE.org are not created with an initial model, but empty. Any UI
allowing to instantiate a model as an instance, or any functionality
instantiating a sub-process instance, therefore in the next step has to load a
process model (through the \ici{} control interface, which in turn triggers events
being sent out through the \idsi{} data stream interface). In ready state (a stopped
state), changes to all aspects of an instance are possible: the instance
model(description), dataelements, endpoints, attributes, as well as the
position in the instance model (description) that the execution should start
from.

From there (1) a UI managing the instance, or (2) a functionality instantiating
a sub-process instance can trigger a transition to state:

\begin{itemize}

  \item \textbf{Running}: The instance is executed, activities are enacted.

  \item \textbf{Abandoned}: A manually set state (without proper execution) signifying
  that the instance is no longer able to run. For example, external services
  connected through the \idsi{} data stream interface might have prohibited the proper
  loading of a process model into the instance, thus rendering the instance
  unusable. This state is final, and can not be left. No further changes to the
  instance are allowed.

\end{itemize}

If the execution of a process instance is successful (without an error
occurring), the instance will transition to state \textbf{Finished}. This state is
final, and can not be left. No further changes to the instance are allowed. The
state finished can not be voted on, and can not be set through the \ici{} control
interface.

If an error occurs during the execution of a process instance, the instance
transitions to the state \textbf{Stopping}. This set can also be triggered for running
instances at any time through the \ici{} control interface, and can be voted on by
external interfaces through subscriptions to the \idsi{} data stream interface.

The state stopping is an intermediate state to give functionalities the chance
to go into a consistent state. Synchronous activities in parallel branches need
still be able to collect responses from invoked functionalities. As soon as all
functionalities have successfully returned values the instance state will
transition to \textbf{Stopped}. Asynchronous activities do not contribute to delays. In
stopped state callbacks will be suspended, when the state changes back to
\textbf{Running}, callbacks are again accepted for an activity. Thus, (1) synchronous
activities have to return before stopped state, (2) asynchronous activities can
be suspended.

The \textbf{Purged} state is only reachable from \textbf{Running} and \textbf{Abandoned}. While for all
other states, the instance can be inspected through the PE, after purging only
logs created through the \idsi{} data stream interface continue to exists.

\subsection{Activity Lifecycle}
\label{sec:al}

\begin{figure}
  \centering
  \includegraphics[width=0.9\textwidth]{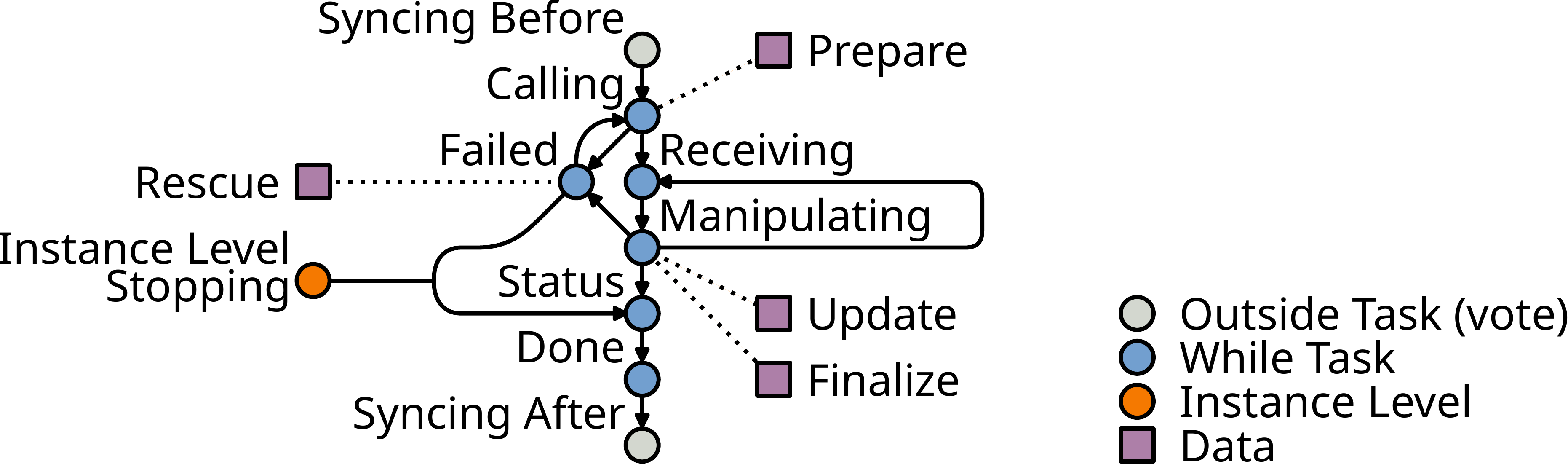}
  \caption{CPEE.org Activity Lifecycle}
  \label{fig:lifecycle-activity}
\end{figure}

Whenever the PE enacts an activity, the activity enactment transitions through
a set of states (see Fig. \ref{fig:lifecycle-activity}), which also results in events being sent to
external services subscribed to the \textbf{activity} topic.

\textbf{Syncing Before} and \textbf{Syncing After} are votes, thus external services can prohibit
or delay the enactment of an activity (cmp.~\cite{mangler_rule-based_2011}).
Both of these states are not part of the formal enactment of the activity but
signify before and after enactment.

The \textbf{Calling} state is the first state of the enactment. It signals that input
data is sent to the functionality implementing a certain task as part of the
enactment of an activity. Before actually invoking functionality a \textbf{Prepare}
script can be used to prepare the input data. Changes to the instance context
(dataelements) made in this script are not permanent and only exist in the
scope of a certain activity.

It the functionality responds, and data is received, the activity transitions
to the \textbf{Receiving} state. Receiving data can happen as part of a
synchronous or asynchronous interaction between an activity and the
functionality it invokes.  Depending on the amount of data the receiving phase
takes a certain amount of time. For synchronous or asynchronous then the state
transitions to \textbf{Manipulating}, so either \textbf{Finalize} or
\textbf{Update} scripts are invoked. \textbf{Update} is called in case of an
asynchronous update interaction (see above), \textbf{Finalize} is called in all
other cases. Both scripts have full access to the received data, as well as to
the instance context (dataelements) and can modify it permanently. In case of
the asynchronous update pattern the state may again switch to
\textbf{Receiving}.

If \textbf{Calling} or \textbf{Manipulating} fails (either by a functionality not available, or a
response signifying some errors, or a update/finalize script having a syntax
error), the activity transitions into the Failed state. In \textbf{Failed} state a
script \textbf{Rescue} can clean up the instance context (dataelements), or set a
special process status to tell the PE if it should retry invoking (\textbf{Calling}) the
functionality, just ignore the error, or transition the instance to state
\textbf{Stopping}. The activity then transitions to state \textbf{Status}, and subsequently to
\textbf{Done}.

\subsection{Task Lifecycle}
\label{sec:tl}

Each task may have its own internal lifecycle, implemented in the functionality
invoke by an activity. This lifecycle can be either hidden from the PE, or made
transparent through a \textbf{CPEE-EVENT} response from the functionality through the
\ioi{} operation interface (see above). If the internal lifecycle is exposed, it will
be sent out through the \idsi{} data stream interface to all external services
subscribed to the topic \textbf{task}.

\begin{figure}
  \centering
  \includegraphics[width=0.8\textwidth]{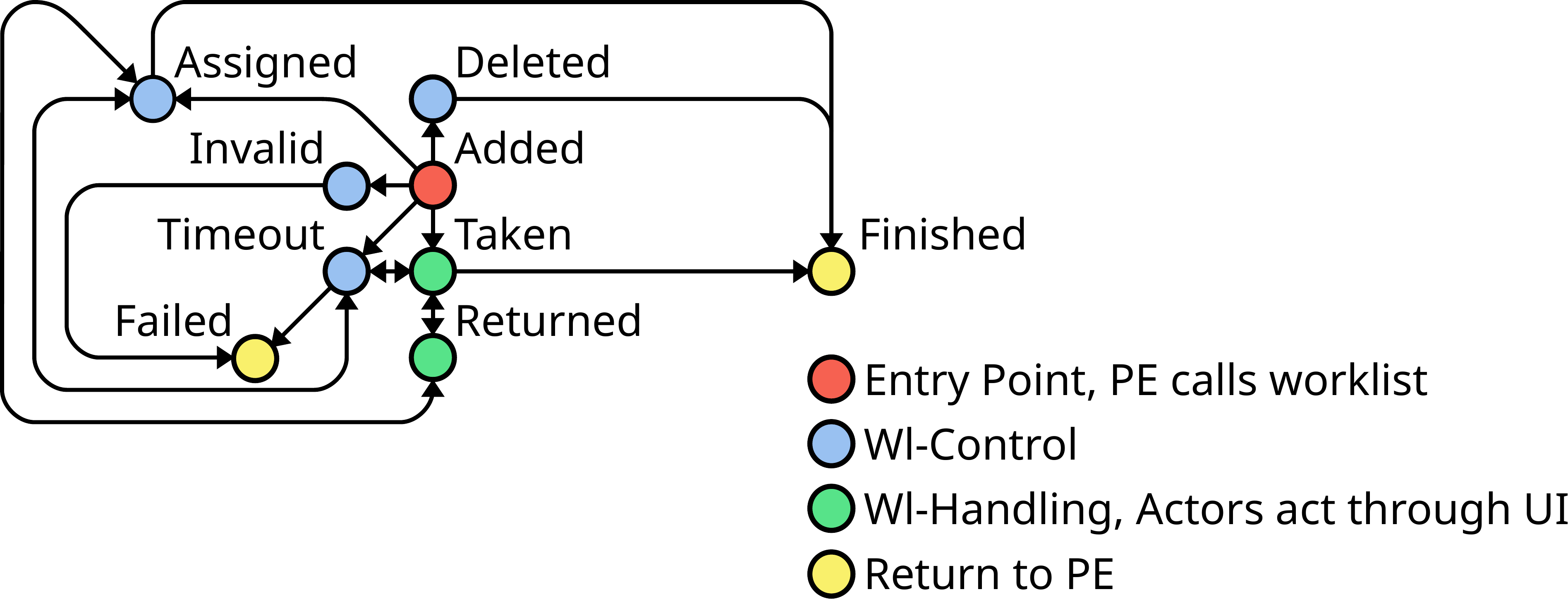}
  \caption{CPEE.org Worklist Lifecycle}
  \label{fig:lifecycle-worklist}
\end{figure}

Fig. \ref{fig:lifecycle-worklist} depicts a potential lifecycle for a human
task, although other human tasks might implement different lifecycles. The
human task in this particular case is implemented as a worklist functionality.
Each enacted activity can pass information about a task to the worklist
functionality, which then coordinates humans to work on the tasks stemming from
different activities in parallel branches, and different instances.

The lifecycle depicted in Fig. \ref{fig:lifecycle-worklist}, depicts the possible states of one task.
Whenever the worklist functionality is invoked as part of the enactment of an
activity, the \textbf{Added} state is reached. As part of the internal functionality of
the worklist the \textbf{Deleted} intermediate state might be triggered (e.g., when the
task is a duplicate), leading to the state \textbf{Finished}.

Alternatively the \textbf{Invalid} state might be reached, e.g., if there is no suitable
human worker being able to work on the task (e.g. because all workers are
unavailable due to illness), leading to a \textbf{Failed} state.

Another possibility is triggering of state \textbf{Timeout} if a supplied deadline has
passed, leading again to the \textbf{Failed} state.

The \textbf{Assigned} state is a special state that can be reached for certain classes of
worklists that automatically assign tasks to humans:

\begin{itemize}

  \item Round Robin worklist: work is assigned to a set of humans (e.g., sharing
  a common role) in round robin fashion. The first task is assigned to the first
  human, the second task to the second human, and so on. When all humans have a
  task, the first human is again assigned a task.

  \item Workload worklists: a random human belonging to a group (e.g., sharing a
  common role) with the lowest number of tasks is assigned a task.

  \item Skill based worklists: the human with the best set of skills matching the
  task description is assigned the task.

\end{itemize}

This state can be reached from \textbf{Added}, as well well as \textbf{Timeout} (e.g., when a
deadline is passed, the task is reassigned to a different human) states. This
state can result in \textbf{Finished} state. Furthermore humans can signal that they can
not do the task resulting in the state \textbf{Returned}. From there the task can be
reassigned to a different human, resulting again in state \textbf{Assigned}.

An altogether different class of worklist is described by the remaining states.
The \textbf{Taken} state can only occur in a worklist where tasks are not automatically
assigned, but instead actively reserved by a human from a list of available
tasks. Each task is typically visible to a group of humans sharing a common
role. Taken tasks are no longer visible to other humans in that group. Taken
tasks can either be \textbf{Finished}, \textbf{Returned} to the list of tasks for other humans in
the common group to be reserved, or automatically \textbf{Assigned} to a human as
described above.

\textbf{Failed} and \textbf{Finished} are the two final states reachable for a task. \textbf{Taken} and
\textbf{Returned} are states triggered by human action, while all other states are
typically the result of worklist internal mechanisms.

Activities invoke Worklists in an asynchronous manner, final responses occur
when \textbf{Failed} and \textbf{Finished} states are reached. All other state changes might lead
to intermediate responses (asynchronous update, see above), and thus to events
send to all external services subscribed through the \idsi{} data stream interface.

\section{Components}
\label{sec:comp}

In order to assemble a service-oriented BPM like CPEE.org, the interfaces
presented above can be used to create and connect a set of components, allowing
for managing and operating a highly-scalable system.

\subsection{Control Interface}
\label{sec:ci}

Connected to the control interface are four main components.

\begin{figure}
  \centering
  \includegraphics[width=1.7\textwidth,angle=270]{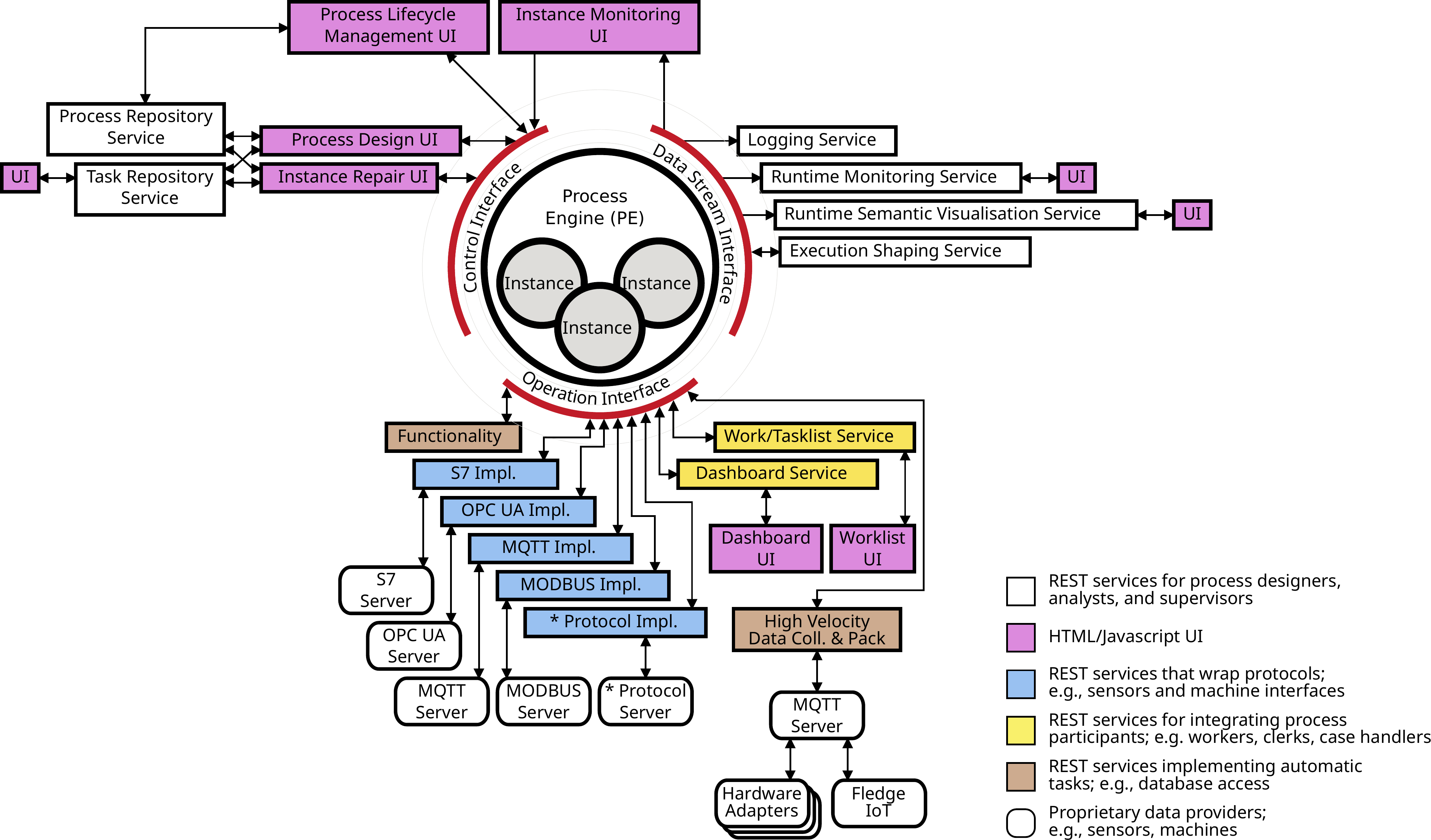}
  \caption{CPEE.org Components And How They Use The Interfaces}
  \label{fig:components}
\end{figure}

\begin{figure}
  \centering
  \includegraphics[width=0.7\textwidth]{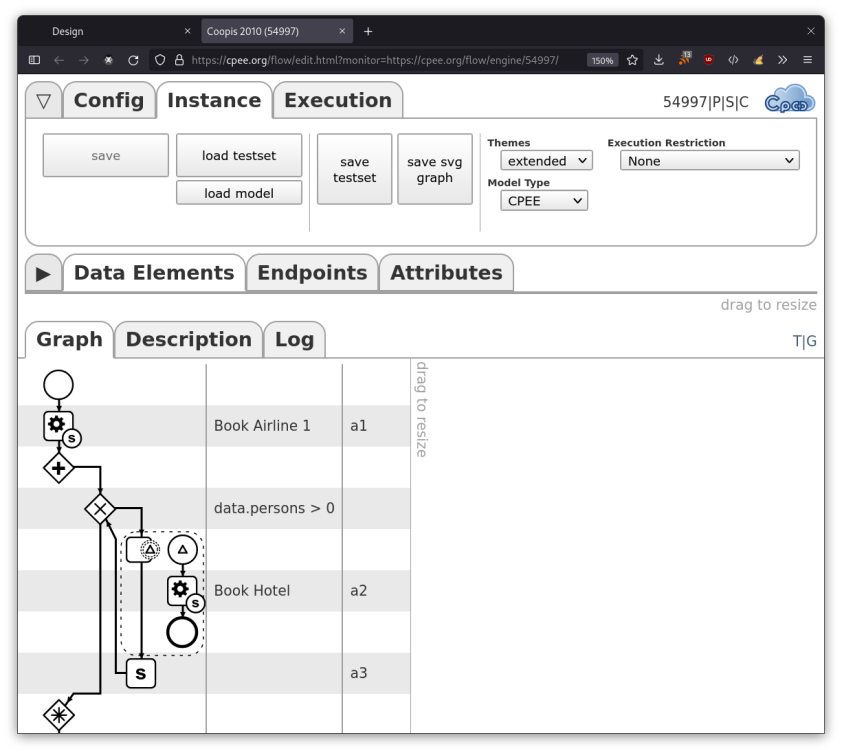}
  \caption{CPEE.org Process Design UI}
  \label{fig:pdui}
\end{figure}

\begin{figure}
  \centering
  \includegraphics[width=1.0\textwidth]{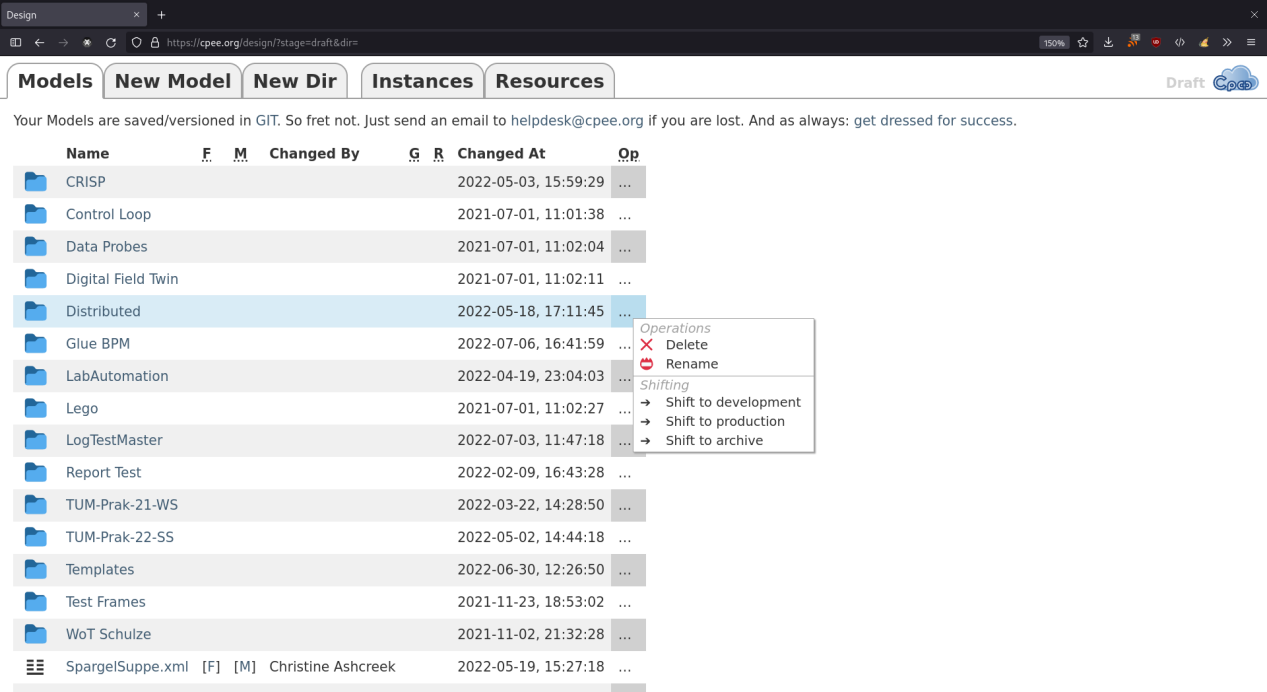}
  \caption{CPEE.org Shifting Between Lifecycles}
  \label{fig:shifting}
\end{figure}

The \textbf{Process Design UI (PDUI)} creates process models, by using the task
repository (simple list of available endpoints), as shown in Fig. \ref{fig:pdui}. The PDUI
is a simple HTML/JavaScript SPA (single-page-application). It is connected to
the \ici{} \textbf{control interface} to allow for testing the models on-the fly (by creating
new instances). Whenever process models reach a certain maturity, they can be
saved in the \textbf{Process Repository (PR)}. The PR, just acts as a storage front-end
which versions each saved process model in arbitrary GIT repositories, which is
important to comprehend changes, and cooperative work on models. Versions are
created whenever a user saves the model into the PR (see Fig. \ref{fig:pdui}, ``Save'' top
left).

The stored and versioned process models can be managed through the \textbf{Process
Lifecycle Management UI (PLMUI)}~\cite{mangler_centuriowork_2019}. The PLMUI
allows to manage the lifecycle of models. Each model can be either a graphical
design draft (no endpoints), under development (not yet fully functional and
tested), in production, or at its EOL in the archive (see Fig.
\ref{fig:shifting}). Each model, or each folder of models can be shifted
between these four lifecycle stages, in the top right of Fig.  8, it is
possible to switch which lifecycle is currently displayed. While these four
lifecycle stages are typical in Software Engineering, it is possible to
configure the PDUI for additional lifecycle stages to match different, more
extensive, or more basic development styles.

The PDUI (see Fig. \ref{fig:pdui}) itself is a designed as a cooperative editor, so when
multiple people work on the same model, all edits are directly shown in all
browsers currently viewing the model.  This cooperative editing is realized
through a SSE (server side events) subscription the \idsi{} data stream interface
(connection not shown in Fig. \ref{fig:components} for simplicity).

The \textbf{Instance Repair UI (IRUI)}, is very similar to the PDUI. It works
on single instances, which are in state stopped (see Sec. \ref{sec:il}
``Instance Lifecycle'' above).  Firstly, whenever changes are made, a user has
the option to save the changes to the process model, for later instances of the
process to include the fix.  Secondly, the user can also apply the fixes to
other running instances, which have not yet reached progressed to the point of
the fix.

\begin{figure}
  \centering
  \includegraphics[width=1.0\textwidth]{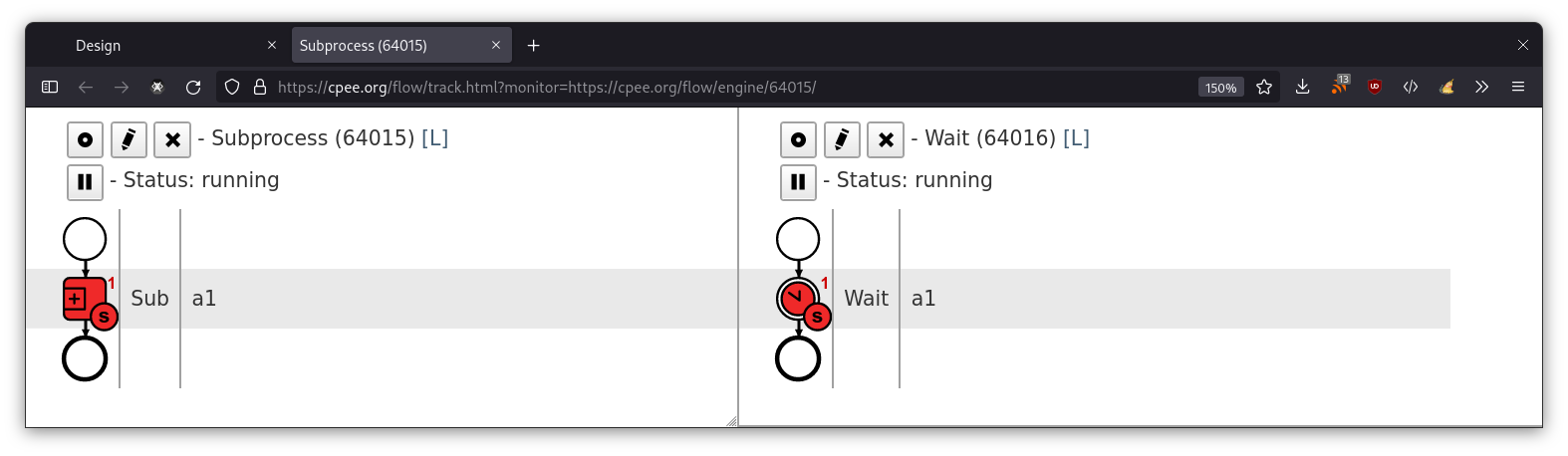}
  \caption{CPEE.org Instance Monitoring UI}
  \label{fig:instance-monitoring}
\end{figure}

The \textbf{Instance Monitoring UI (IMUI)} (see Fig. \ref{fig:instance-monitoring}) shows a life view of how
instances are executed (red task is currently executed task). If an instance
spawns one or many sub-process, they are shown to the left of the instance. For
each instance, the IMUI offers controls to change the state (e.g. stop, start),
change to edit mode (i.e., IRUI), or hide an instance from view. The IRUI is
again an HTML/JavaScript SPA, which is subscribed to the \idsi{} data stream interface,
to receive information about which task is currently executed (red task), and
the current state of each instance. It is also subscribed to the task topic to
receive instantiation events, in order to show sub-processes.

\subsection{Data Stream Interface}
\label{sec:dsi}

Exclusively connected to the \idsi{} data stream interface are four main components.

The \textbf{Logging Service} has no UI. Its purpose is to subscribe to a set of events,
in order to store an XES file on disk. The XES files are linked in the IRUI
(see Fig. \ref{fig:pdui}, Log UI Element, in the main/bottom right).

\begin{figure}
  \centering
  \includegraphics[width=1.0\textwidth]{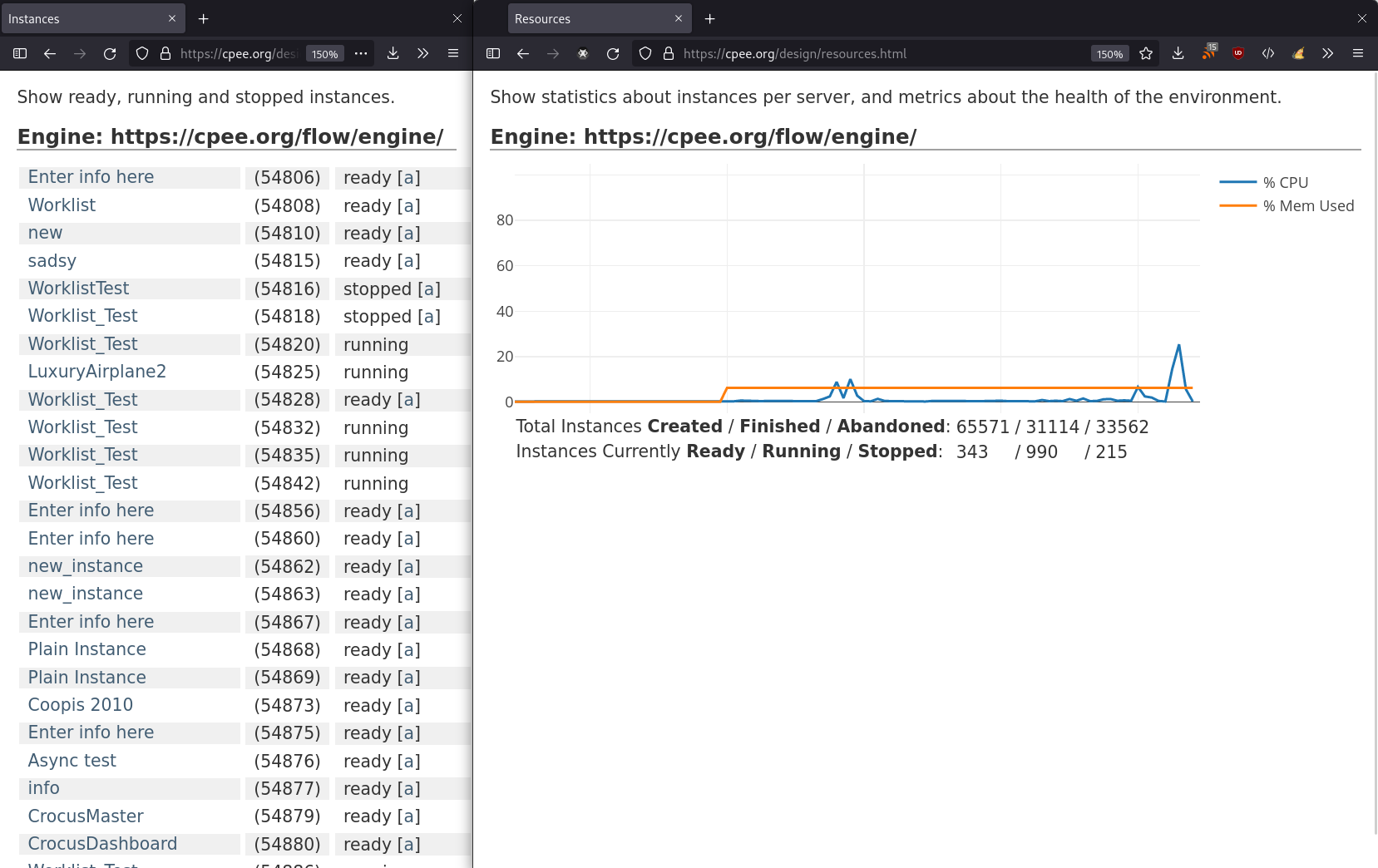}
  \caption{CPEE.org Runtime Monitoring}
  \label{fig:runtime-monitoring}
\end{figure}

The \textbf{Runtime Monitoring Service + UI (RMUI)} (see Fig. \ref{fig:runtime-monitoring}) consume events from
the \idsi{} data stream interface to provide information:

\begin{itemize}

  \item Running Instances, their state (running, ready, stopped), as well as the memory usage per instance (left)

  \item Statistics about the overall memory and CPU usage (right).

  \item Statistics about total instances, as well as currently active instances (right).

\end{itemize}

Please note that the URL of the engine is very prominently shown, as the RMUI
can be subscribed to multiple engines, which either together format a
load-balanced cluster of engines or are unrelated. The functionality shown in
Fig. \ref{fig:runtime-monitoring} is achieved by the following subscription:

\begin{itemize}

  \item topic \textbf{state}, event \textbf{change}: monitor the creation of instances (ready), as
  well as their full lifecycle as described in the Sec. \ref{sec:il}.

  \item topic \textbf{task}, event \textbf{instantiation}: monitor the creation of sub-process
  instances. While \textbf{state/change} only provides information about the existence of
  an instance, this adds information about their parent/child relationship.

  \item topic \textbf{status}, event \textbf{resource\_utilization}: monitor the memory and CPU
  usage for instances.

\end{itemize}

A \textbf{Runtime Semantic Visualization Service + UI (RSVUI)}, is in contrast to the
RMUI intended to subscribe mostly to topic \textbf{dataelements}, to monitor the
data-flow in process instances, and topic \textbf{activity} to monitor the duration of
individual tasks. An RSVUI is a custom \textbf{Key Performance Indicator (KPI)}
monitoring and visualization service. For example, when a service is about
production of parts, the number of parts, cycle times (time to produce one
part, which might be the result of multiple activities enacted in a loop) or
overall equipment efficiency (OEE) can be displayed. This can be realized in
two ways:

\begin{itemize}

  \item Bad solution: write a custom service, the hard-codes the meaning of
  certain data-element and tasks, in order to find and display the KPIs Whenever
  the process model changes, it has to be checked if the RSVUI has to be changed
  as well, as it might breach when new activities are added, or the data flow
  changes.

  \item Good solution: annotate the BPMN with semantic information about how to
  extract the KPIs. Thus the RSVUI will be more generic and can, whenever events
  are received, inspect the corresponding BPMN for data extraction,
  transformation and display (ETD) information.

\end{itemize}

Both solutions can be observed in practical applications. While the first
solution is sometimes preferred for less implementation overhead, the second
solution is always better given that a suitably powerful semantic annotation,
mechanism exists. For CPEE.org various aspect of such a BPMN extension are
still subject of on-going research~\cite{ehrendorfer_sensor_2021}, for other BPMN
editors, such research to the best of our knowledge is not easily possible or
foreseen.

An \textbf{Execution Shaping Service (ES)} is a special component that subscribes to
arbitrary events and votes from the \idsi{} data stream interface, and \textbf{enacts actions
through votes}  as described in Sec. \ref{sec:dsi} ``Data Stream Interface''. Examples
for such services are:

\begin{itemize}

  \item Runtime compliance checking: directly reacting when compliance violations
  are detected. Simple cases might include stopping an instance, and notifying
  responsible actors, complex cases might include the automatic modification of
  responsible actors in the instance process model to fix the compliance
  violation.

  \item Self-healing: in case of errors occurring with endpoints (e.g.,
  machines), fix the instance by changing endpoints, or triggering compensation.

  \item Load-Balancing: at runtime change endpoints to select resources
  (endpoints) with the least workload.

\end{itemize}

Many other applications exist, as for RSVUI these applications might be very
domain specific and might require additional information in the process model
to keep the ES generic enough to not break its functionality when process
models evolve.

\subsection{Operation Interface}
\label{sec:oi}

Components using the \ioi{} operation interface should not use any other interfaces,
as this will negative effects on the security of the overall system. Separation
the enactment of activities, or rather the functionalities they are linked to,
will guarantee that everything can be properly tracked and observed, and no
behaviour can be hidden.

For CPEE.org currently all components connected to the operation interface are
realized as REST services, although this is not a requirement, as the engine
supports pluggable \ioi{} operation interfaces that could support arbitrary protocols.

For CPEE.org all REST services follow the synchronous, asynchronous, or
asynchronous update patterns, as described in Sec. \ref{sec:oi}.

The services connected to the \ioi{} operation interface fall into four generic
groups:

\begin{itemize}

  \item \textbf{Basic Functionality}: unspecified functionalities that preform automatic
  tasks such as extracting data from a database, or extracting data from CRM.
  They are black boxes, and implement a specific interface.

  \item \textbf{Protocol Proxies}: a class of services that wrap custom, often
  proprietary protocols. Currently CPEE.org supports S7, OPCUA, MQTT and
  MODBUS, which proved sufficient for many industrial applications. As these
  protocols might have very different communication patterns which might solely
  rely on pushing messages, the proxy service mostly use the asynchronous
  pattern to interface with the process engine.

  \item \textbf{User Integration}: User integration is again a form of proxy service,
  but with the goal of integrating users into an enacted. For this a
  Work-/Tasklist or Dashboard Service has to utilize the HTTP headers as
  described in Sec. \ref{sec:oi} ``Operation Interface'', store callback information and
  parameters in provided parameters, and tell the process engine that
  asynchronous pattern is to be used. A separate UI, solely used by Process
  Participants then utilized the provided parameters to present an UI, as
  discussed above. CPEE.org provides both, a worklist as well as a dashboard
  component to build interactive user interfaces. While in office automation
  worklists are more common, on the shop-floor dashboards are more prevalent.

  \item \textbf{High velocity data collection and packaging}: In IoT
  environments such as shop-floors, typically two kinds of components exist:
  Sensors and Actuators. Actuators can be triggered to start some sort of
  operation. They are typically directly represented as activities in process
  models, and typically return a result describing success/error of an
  actuation. Sensors on the other hand observe various properties of the
  shop-floor. They might range from something as simple as measuring
  temperature and humidity, to monitoring all conditions inside a machine, such
  as the power consumption of individual motors, or the position of various
  axes of a lathe. There are two different scenarios that might occur:

  \begin{itemize}

    \item Sensor information is related to one particular activity. E.g., for
    an activity ``Machine Part'' all sensor information regarding this machining
    operation can be collected, and directly attached to the task.

    \item Sensor information is continuous and not related to on particular
    activity, but rather to a group of activities, an instance, or even a group
    of instances. Thus the information can not be attached to single activities
    but to a higher plane of structure.

  \end{itemize}

\end{itemize}

\textbf{User Integration} is the most prominent use-case for utilizing the \ioi{} operation
interface. Traditionally work-/tasklists where integrated into monolithic BPMs,
instead of being loosely coupled with it through a common interface for all
services. This can be also seen when looking, e.g., at the XES standard for
storing log information. Its lifecycle extension is mixing the lifecycle for
activities (see above) with the Lifecycle for tasks (see above), because in the
traditional view there was no separation between these. The traditional view
does not consider the different/specialized work-/tasklists could have
different and much more fine-grained lifecycles.

Traditional work-/tasklist (being integrated into the BPM) also assume that
they have access to the full process context, i.e. all dataelements, attributes
and other internal information. For CPEE.org, with its focus on modularity,
loose coupling, and strict separation of concerns, this is not true. Utilizing
the \ioi{} operation interface induces that all information required for a
work-/tasklist do do its job is passed to the functionality that implements it.
Information required by a tasklist might include:

\begin{enumerate}

  \item[(A)] Which user/role is to work on a task.

  \item[(B)] Which organization structure is to be used to select users based
  on roles.

  \item[(C)] UI/Form which should be shown when working on the task.

  \item[(D)] Task specific information that has to be known to the user working
  on the task in order to be able to do it.

  \item[(E)] Deadlines.

\end{enumerate}

A \textbf{work-/tasklist} should return (at least) at least (1) the result of the work
(e.g., a document or success notification), and (2) which user(s) worked on the
task in order to be able to do compliance checking.

It is not wise to handle Separation of Duty (SoD) / Binding of Duty (BoD)
internally in a work-/tasklist, but to escalate it to the level of the process.
This should be realized by treating the passed information (a) (see enumeration
above) as follows:

\begin{itemize}

  \item Separation of Duty (SoD): in addition to the role (list of people who can
  do a task), pass information about a user / users from this list, who are NOT
  allowed to do this task. The not-allowed-users can be collected from the return
  of previous calls to work-/tasklists which are part of the SoD logic.

  \item Binding of Duty (BoD): instead of passing a role  (list of people who can
  do a task), only the user that preciously worked on tasks which are part of the
  BoD logic.

\end{itemize}

By implementing this on the level of the process, two main advantages:

\begin{itemize}

  \item Simple Compliance Checking: Compliance checking can be realized at the
  process level, instead of being required to access information from private
  logs of the work-/tasklist service.

  \item Centralized Configuration: All  configuration is part of the process
  model, instead of being hidden inside functionality. Different
  work-/tasklists can be mixed, without the requirement of accessing common
  configuration information.

\end{itemize}

\textbf{Dashboards} are different from work-/tasklists, as they are either (1)
read-only, or (2) bound to a physical location. Because of this, they might be
access restricted, but there is no need to specify user/role or organizational
information. They also typically only show one thing. So when triggering a
dashboard either (1) the information shown is replaced, or (2) added to be
shown simultaneously in an other part of the dashboard.

High velocity data collection and packaging realizes a special service which
can be used in two ways:

\begin{itemize}

  \item A process engine (PE) can ask for packaged data from not one or a group
  of sensors. In this case the process model contains instructions to
  explicitly collect sensor information.

  \item A process engine (PE) implements that in parallel to the execution of
  instances or the enactment of activities sensor information is collection. In
  this case the process models contains additional information to collect
  information while an instance is executed or while a group of activities is
  enacted.

\end{itemize}

CPEE.org implements both mechanisms, both through normal BPMN and BPMN IoT
extensions.

\section{Highly Scalable Architecture}
\label{sec:arch}

\begin{figure}
  \centering
  \includegraphics[width=1.0\textwidth]{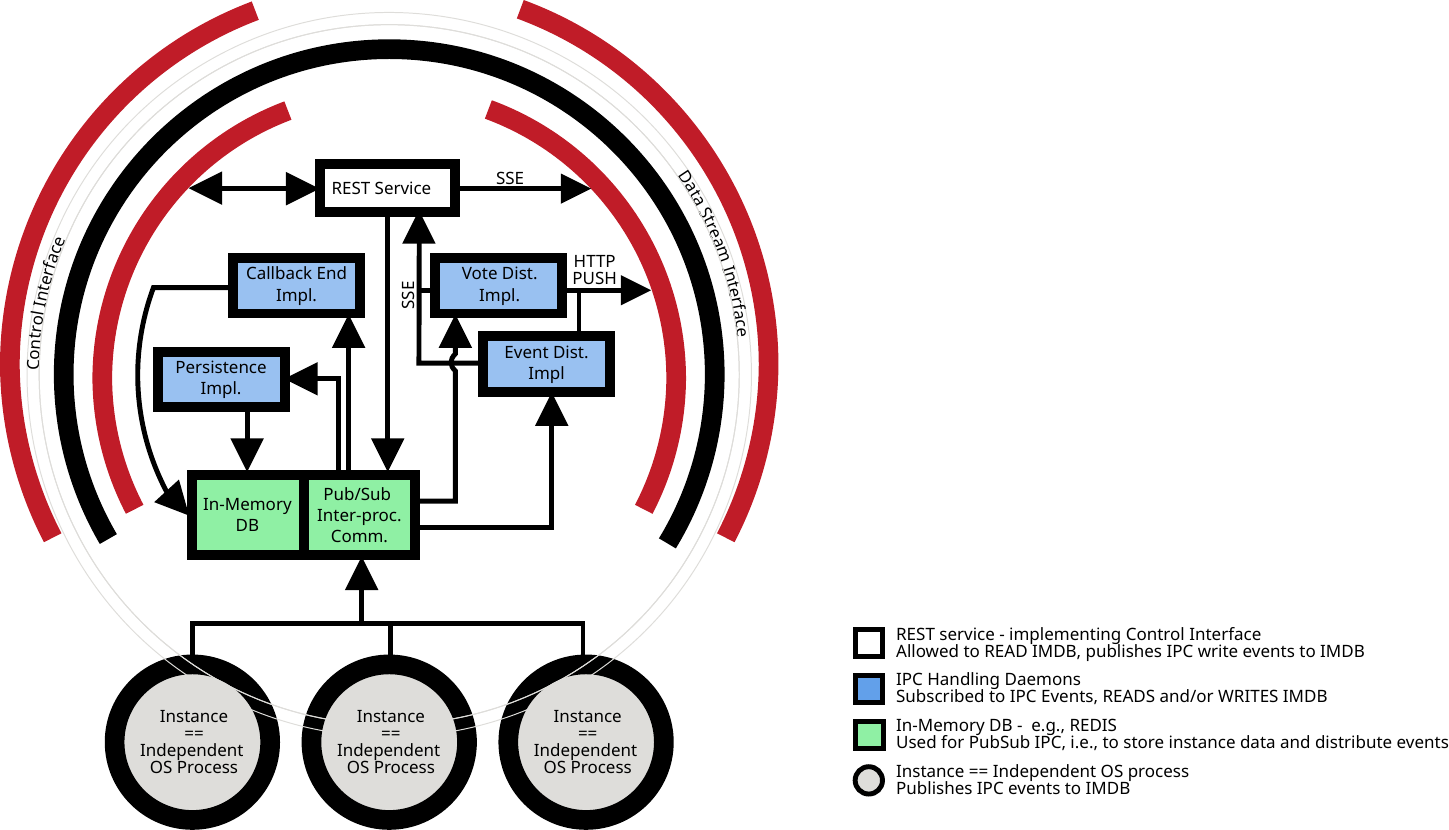}
  \caption{CPEE.org Highly Scalable Service Oriented Architecture - Internal Message Routing}
  \label{fig:scale}
\end{figure}

CPEE.org is not only modular by providing a set of interfaces to the outside,
but it it is highly scalable by internally also being based on a set of
services. This allows to distribute one CPEE.org over multiple nodes (scale-out
architecture) in a number of different communications, best suiting the needs
of a wide range of application scenarios. As depicted in Fig. \ref{fig:scale},
the first important part is: \textbf{Each instance is a separate service},
communicating with the rest of the PE through an IPC mechanisms with
Publish/Subscribe functionality, connected to an in-memory data-base to store
the internal state of an instance execution. This introduces the following
properties:

\begin{itemize}

  \item Instances can be deployed to different nodes for maximum scalability.

  \item Instances can be deployed into separate containers for improved security properties.

  \item Instances are managed by the underlying OS as processes, meaning any number of CPU cores is transparently used.

  \item Instances can be monitored through standard monitoring facilities, their CPU and memory usage is always separately available.

  \item Instances can be restricted with separate CPU / Memory quotas through standard OS mechanisms.

\end{itemize}

\textbf{Each instance implements the \ioi{} operation interface}. Each instance can be realized
in two different ways:

\begin{itemize}

  \item By having an interpreter read and execute the statements in a BPMN.

  \item By translating/transpiling the BPMN into a native language and then
  compiling/executing the result.

\end{itemize}

CPEE.org uses the second mechanism as it provides higher performance and lower
overhead. In general it has to noted that realizing instances as standalone
services has also some drawbacks:

\begin{itemize}

  \item The necessity of using IPC introduces some serious overhead, when
  compared to having a multitude of instances being executed (interpreted)
  inside a monolithic process engine. We think the possibility of scaling
  mitigates this disadvantage.

  \item The memory overhead can be considerable, as each instance potentially
  has to carry and run its own BPMN interpreter. CPEE.org avoids this by
  employing the transpilation mechanism.

\end{itemize}

CPEE.org employs Redis\footnote{\url{https://redis.io/}} as the in-memory database. Any change to dataelements,
endpoints, or attributes, has to be made available to the Pub/Sub mechanism by
the instance. Furthermore information about which task is currently executed
(including lifecycle information) has to be made available as well. In fact,
all information available as an event, as described in Sec. \ref{sec:lifecycles}) has
to be constantly sent while each instance is running.

The remaining components inside the PE are subscribed to this event-stream and
act on it as follows:

\begin{itemize}

  \item The \textbf{Persistence Implementation (PI)} saves the information in the
  in-memory database.

  \item The \textbf{Event Distribution Implementation (EDI)} directly sends relevant
  information to subscribed services, according to the logic described in Sec.
  \ref{sec:dsi}. While HTTP based push messages can be distributed (fire and
  forget) directly to subscribers, Server Sent Event (SSE) subscribers have to
  be handled differently (see below).

  \item The \textbf{Vote Distribution Implementation (VDI)} does the same as the EDI but
  for votes. This component also handles the answers to votes, which can only
  arrive through HTTP (see below).

  \item The \textbf{Callback End Implementation (CEI)} which is responsible for cleaning
  up the database after a response to a vote or an asynchronous call.

\end{itemize}

The final and most important service is \textbf{REST Service (RS)}  which implements the
\textbf{\ici{} control interface}. All input to the \ici{} control interface again is sent to the
Pub/Sub mechanism, and thus is distributed to the internal components. Each
\textbf{process instance} is subscribed to certain events as well. For example an
stopping state change, triggered through the \ici{} control interface, has to be
received by the respective instance, which then has to stop running: (1) it has
to wait for all synchronous calls though the \ioi{} operation interface to finish, (2)
has to announce state stopped, and (3) has to exit (on the OS level, thus
releasing all memory).

While EDI and VDI actively distribute votes and events over HTTP, the
possibility to subscribe to this through SSE, brings the necessity for the REST
service RS to deliver events as well. Thus the EDI and VDI send special IPC
messages, to which the RS is subscribed, which are then sent through SSE. This
allows for example the \textbf{Instance Monitoring UI} component (see section above),
which is just HTML and JavaScript to receive the necessary information to
update its UI.

\begin{figure}
  \centering
  \includegraphics[width=0.9\textwidth]{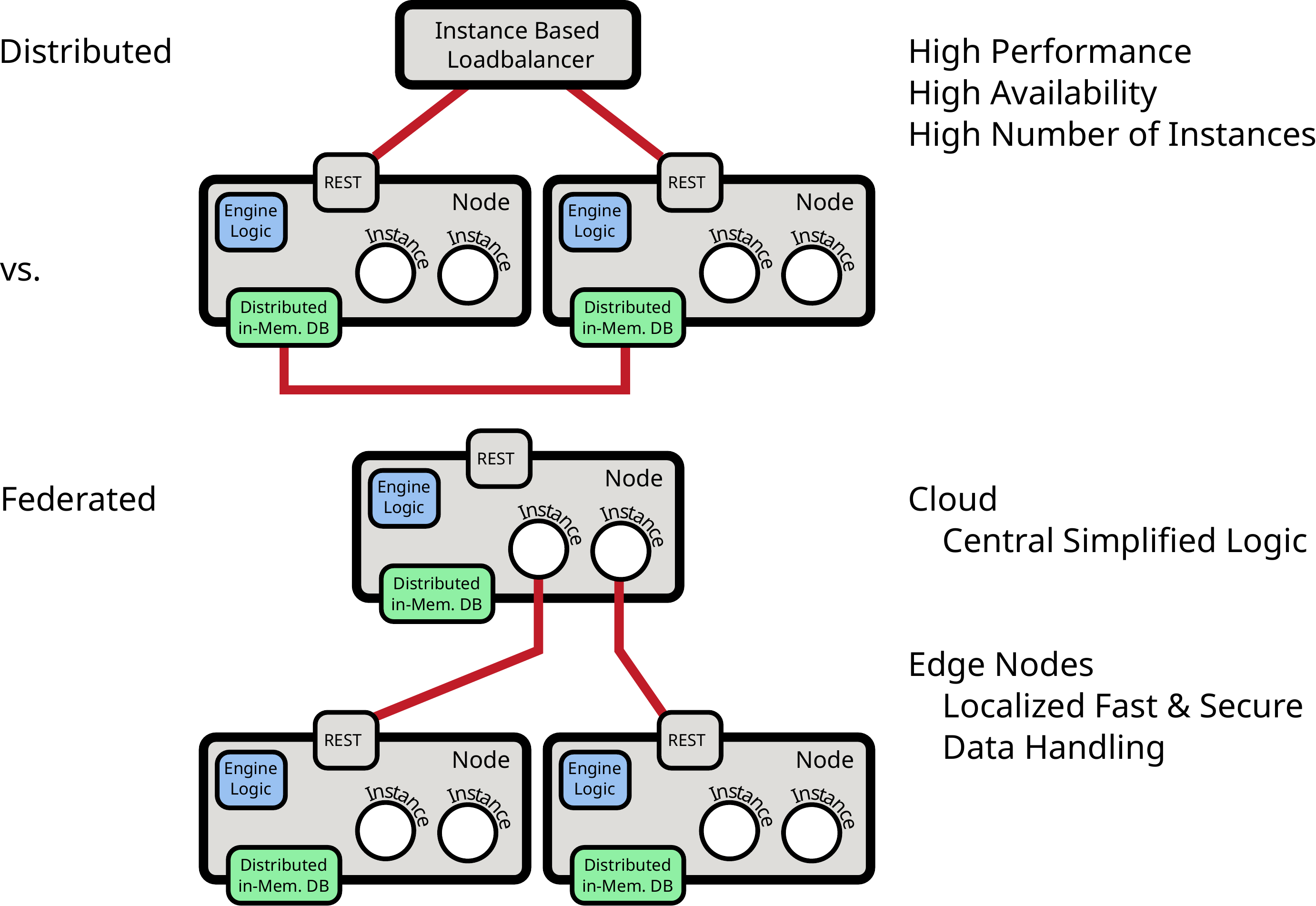}
  \caption{CPEE.org Distributed Vs. Federated}
  \label{fig:dist}
\end{figure}

With these internal services, the PE can be set up in many possible ways, some
of which are depicted in Fig. \ref{fig:dist}. In the \textbf{Distributed Scenario}:

\begin{itemize}

  \item Each node has its own Distributed in-memory database, but they are
  operating as one cluster. This is for example supported by Redis as used in
  CPEE.org, but also supported by others.

  \item Instances run on nodes.

  \item All other services, including the REST service are available on each
  node.

  \item A centralized load-balancer distributes \textbf{\ici{} control interface} HTTP traffic
  based on the instance id. E.g., for two nodes all even instances are hosted
  on node one, all odd instances on node 2.

\end{itemize}

Of course different more complicated load-balancing mechanisms can be realized
easily. Another possibility is to not host the instances on the same node but
to distribute them to separate nodes. Separate nodes for EDI/VDI, which are
easily the most taxing services, are possible as well.

Another important scenario is the \textbf{Federated Scenario}. For IoT/Edge use-cases
different Process Engines (PE) can run on different nodes. As the creation of
sub-process instances is realized through a special \textbf{\ioi{} operation interface}
component, federated PEs can be used like normal services.

\section{Conclusion}
\label{sec:conclusion}

CPEE.org realizes a Process Engine, which goes beyond the state-of-the of both,
industrially and scientifically available offerings. Its core and many
components are open-source\footnote{\url{https://github.com/etm/}}, actively
maintained and constantly extended. Its no-compromise architecture makes it
particularly well suited for taxing industrial applications. It is also well
suited for University teaching, due to its robustness (separate instances) and
security (instances runnable in containers).

Due to its modularity, while maintaining three simple and streamlined
interfaces, it is very well suited for research. All aspects of BPMN can be
customized through external services, without the necessity of learning any
specific technologies or programming languages.

While more then 15 years old, it maintains a healthy community of developers
and users, both from
industry~\cite{pauker_centuriowork_2018} and
academia.

\interlinepenalty10000
\bibliographystyle{splncs04}
\bibliography{cpee.bib}

\end{document}